\documentclass[final,rnote]{aa}
\usepackage{epsfig} 
\usepackage{epsf}
\usepackage{natbib}
\usepackage{amssymb}
\usepackage{color}
\usepackage[american,english]{babel}

\def\dfnull{\ensuremath{\delta F_0}}
\def\dflim{\ensuremath{\delta F_{\mathrm{lim}}}}
\def\nlim{\ensuremath{N_{\mathrm{lim}}}}
\def\ntot{\ensuremath{N_{\mathrm{tot}}}}
\def\cnfull{\ensuremath{\sigma_0}}
\def\cnlim{\ensuremath{\sigma_{\mathrm{lim}}}}

\begin{document}

\newcommand{\lele}[3]{{#1}\,$\le$\,{#2}\,$\le$\,{#3}}
\title{The impact of main belt asteroids on infrared--submillimetre\\
 photometry and source counts}
\author{ Cs.~Kiss\inst{1} 
  \and A.~P\'al\inst{2}
  \and T.G.~M\"uller\inst{3}
  \and  P.~\'Abrah\'am\inst{1}}
\institute{ Konkoly Observatory of the Hungarian Academy of Sciences, 
    P.O. Box 67, H-1525~Budapest, Hungary
    \and
    Department of Astronomy, E\"otv\"os University, 
    P\'azm\'any P\'eter st. 1/A, 
    H-1117 Budapest, Hungary
    \and
    Max-Planck-Institut f\"ur extraterrestrische Physik, 
    Giessenbachstrasse, D-85748 Garching, Germany
  }
\offprints{Cs.~Kiss, pkisscs@konkoly.hu}
\date{ Received  / Accepted ...}
%
\abstract
{Among the components of the infrared and submillimetre sky background,
the closest layer is the thermal emission of dust particles and minor bodies 
in the Solar System. This contribution is especially important
 for current and future infrared and submillimetre space instruments --
  like those of Spitzer, Akari and Herschel -- and must be characterised 
  by a reliable statistical model.}
{We describe the impact of the thermal emission of main belt asteroids on 
the 5...1000\,$\mu$m photometry and source counts, for the current 
and future spaceborne and ground-based instruments, in general, 
as well as for specific dates and sky positions.}
{We used the statistical asteroid model (SAM)
to calculate the positions of main belt asteroids down to a size 
of 1\,km, and calculated their infrared and submillimetre brightness
using the standard thermal model. Fluctuation powers, confusion noise values 
and number counts were derived from the fluxes of individual asteroids. }
{We have constructed a large database of infrared and submillimetre fluxes for 
SAM asteroids with a temporal resolution of 5 days, covering the time span 
January 1, 2000 -- December 31, 2012. Asteroid fluctuation powers and number counts 
derived from this database can be obtained for a specific observation setup via 
our public web-interface.}
{Current space instruments working in the mid-infrared regime (Akari and Spitzer Space
Telescopes) are affected by asteroid
confusion noise in some specific areas of the sky, while the photometry of space infrared
and submillimetre instruments in the near future (e.g. Herschel and Planck Space 
Observatories) will not be affected by asteroids. Faint main belt asteroids might 
also be responsible for most of the zodiacal emission fluctuations near the ecliptic. }

\authorrunning{Cs. Kiss et al.}
\titlerunning{Impact of asteroids on infrared--submillimetre
 photometry and source counts}	  
	  
\maketitle

\section{Introduction}
\label{sect:intro}
Due to their relatively high apparent brightness at infrared 
wavelength compared e.g.~to Galactic stars, asteroids are among the 
 dominant sources at infrared wavelengths and can seriously affect the 
 infrared and submillimetre photometry and source counts.
Currently, 377\,328 minor planets are 
known\footnote{http://cfa-www.harvard.edu/iau/lists/ArchiveStatistics.html}
(as of July~30, 2007) in our Solar System,
of which about 99\% are located in the main belt.
On the plane of the sky the vast majority of main belt asteroids (MBAs) are 
found at ecliptic latitudes below $20^{\circ}$.
They have sizes between a few ten meters up to about $1000~\mathrm{km}$.
With temperatures between 200 and 300\,K, the asteroids emit
predominantly at thermal wavelengths between $5~\mu\mathrm{m}$ and
the millimetre range. Deep infrared observations close to the
ecliptic will therefore always include some of these moving targets
\citep[e.g.][]{Tedesco+Desert,Meadows2004}.
Such observations also show that only a small fraction of the
existing minor body population is currently known and this population
might cause a non-negligible confusion noise contribution at certain
wavelengths and for specific instruments.
Tedesco \& Des\'ert (2002) measured the number of main belt asteroids
    for the first time in a direct way by using the Infrared 
    Space Observatory (ISO).
    Based on a statistical asteroid model the authors concluded that
    there are about 1.2$\pm$0.5$\times$10$^{6}$ asteroids ($\ge$ 1\,km in
    diameter) in our solar system, twice as many as previously believed.

A set of celestial sources may affect compact source 
observations in two ways:
(1) they contribute to the confusion noise, the uncertainty in
point source photometry due to the fluctuations of the sky background; and
(2) they also appear as individual sources, that may
add 'false' detections to source counts. 

Recently, several authors calculated confusion noise and detection
limits for current/future infrared space missions 
(Spitzer, Akari, Herschel and SPICA).
These papers considered the two major confusion noise components: 
the extragalactic background \citep{Lagache2003,Negrello},
the Galactic cirrus emission \citep{Kiss2005,Jeong2005}, or 
the combination of the two \citep{Jeong2006}. 

It is an important question for infrared space missions
whether faint asteroids close to or below the
detection limit could contribute significantly to the confusion noise
of these instruments or could be present as a significant count of 
contaminating point sources in the field of view. 
To consider these asteroids, a reliable
statistical model is needed.
Recently, \citet{Tedesco2005} presented the "Statistical
Asteroid Model" (hereafter SAM). This model is based on a
population of $\sim1.9\times 10^6$ asteroids obtained from
the complete known asteroid sample (as of 1999), plus
extrapolation of the size-frequency distribution (SFD)
of 15 asteroid dynamical families and three background
populations, to a diameter limit of $1~\mathrm{km}$. 
The validity of the SAM was demonstrated by comparing
SAM predictions with ISO measurements at $12~\mu\mathrm{m}$
\citep{Tedesco+Desert} and Spitzer measurements at the $8$ and
$24~\mu\mathrm{m}$ photometic bands \citep{Meadows2004}.
The asteroid counts from both deep surveys show
good agreement with the SAM predictions.

In this paper we give 
(1) estimates for the impact of MBAs on infrared (IR) to 
submillimetre (especially
spaceborne) measurements, such as asteroid number counts and confusion noise,
for a specific date, sky position, wavelength/instrument; and
(2) observable predictions for the infrared regime which can be 
used to check the parameters of the present model, e.g.~size,
albedo and positional distribution. This is the first study that
investigates the impact of asteroids on infrared and submillimetre photometry 
and source counts in detail, based on a realistic asteroid sample. 


 
\section{Data processing}
\label{sect:dataproc}

\subsection{Position calculation}
\label{sect:poscalc} 

Orbital elements of all SAM asteroids, including the real and predicted 
ones were calculated for different epochs between the time span January~1, 
2000 and December~31, 2012 with an average step size of 5 days. 
This step size is fine 
for yielding a good coverage of the solar elongation in a year. The 
orbital elements were obtained using accurate numerical integrations 
including the effect of all inner and outer planets. From 
the orbital parameters the apparent ecliptic coordinates, distances 
and magnitudes were derived using the spatial coordinates of the Earth 
itself and the absolute magnitudes (which is known from the SAM database). 
Some of the spacecrafts, including the Herschel and Planck 
Space Observatories will be located at the L2-point of the Earth-Moon system.
The difference between the spatial position of the barycentre of 
the Earth-Moon system and the prospective position of the L2-point is 
negligible for the apparent distribution of the asteroids. 
However, the Spitzer Space Telescope is in a significant distance 
from Earth, therefore we performed the position and all subsequent
other calculations for the Spitzer Space Telescope coordinates for a 
limited time span, covering the expected cryogenic lifetime between 
January~1, 2004 and December~31, 2009. The actual positions of the
Spitzer Space Telescope were taken from the NASA/JPL HORIZONS 
system\footnote{http://ssd.jpl.nasa.gov/?horizons}. 

To test the positional accuracy of our integration, we compared 
the observable ephemerides yielded by the
integration and the coordinates returned by the Minor Planet \& Comet
Ephemeris Service of the Harvard-Smithsonian Center for 
Astrophysics\footnote{http://www.cfa.harvard.edu/iau/MPEph/MPEph.html} 
for a couple of known minor and dwarf planets, including Ceres, Pallas, 
Vesta and Astraea. 
The comparison timespan was almost 105 years, resulted by
a backward integration from March 6, 2006 to 1901.0. The differences
between the two sets were always smaller than 0\fdg04 for a specific minor
planet. We have also compared this difference for a timespan of 12 years,
the same as the maximal integration period of the complete simulation. 
In this case the error was definitely smaller than 0\fdg004.

\subsection{Thermal brightness calculation}
\label{sect:tbr}

For each SAM asteroid and for each date thermal fluxes were assigned at 14 
fixed wavelengths \{$\lambda_{0,k}$\}. These wavelengths were chosen 
the cover the 5\,$\mu$m to 1\,mm range in a logarithmically equidistant way. 
If a $\lambda_i$ wavelength was different from \{$\lambda_{0,k}$\}, the
monochromatic flux values were interpolated to the desired $\lambda_i$ 
for each asteroid, individually.
For the brightness calculations we applied the Standard Thermal Model
\citep[STM, see][]{Lebofsky1986}. 
In this model the surface temperature distribution is calculated using the
true heliocentric and geocentric distances. 
The asteroids are described as
smooth, spherical and non-rotating bodies in instantaneous
equilibrium with the solar radiation. No heat conduction into
the surface is considered. The correction for beaming, shape and
conductivity effects is done via the $\eta$-parameter with a
value of $\eta=0.756$. Furthermore, the flux at non-zero solar
phase angles is obtained by applying an empirical phase correction
of 0.01\,mag\,deg$^{-1}$ to the flux calculated at opposition.
The STM has clear limitations with respect to
flux accuracy \citep[e.g.][]{Muller2004} or for modelling
of minor bodies outside the main belt \citep[e.g.][]{Harris1998}, but
highly accurate flux predictions are not crucial for our goals.

%

\subsection{Number counts, fluctuation power and confusion noise}
\label{sect:powercalc}


\noindent{\bf Number counts}: 
Two kinds of number count quantities are calculated: 
(1) $\ntot(\lambda_i)$, 
the total count of asteroids in the counting cell, 
normalized by the solid angle of
the counting cell $\Omega_c$ ($\mathrm{sr}^{-1}$); and
(2) $\nlim(\lambda_i,S_{\mathrm{lim}}$), the
count of asteroids above the detection limit $S_{\mathrm{lim}}$ in 
a particular counting cell, normalized by the solid angle of
the counting cell $\Omega_c$ ($\mathrm{sr}^{-1}$). 

\smallskip
\noindent{\bf Fluctuation powers}: 
The \emph{full fluctuation power} 
\citep[see][for an introduction]{Lagache2003} 
is calculated from the 'observed' distribution of all asteroids in that
specific cell, for a specific $\lambda_i$ wavelength:
\begin{equation}
\delta F_0 (\lambda_i) = \left(\frac{1}{\Omega_c}\right)
\sum\limits_{j} S_j^2(\lambda_i)
\label{eq:dF0}
\end{equation}   
In this case the summation runs over \emph{all} of the asteroids in the 
counting cell ($\mathrm{sr}^{-1}$). 

The fluctuation power due to non-detectable 
asteroids, \dflim, can be calculated for a specific instrument 
in a similar way as \dfnull, but in this case only 
asteroids below the detection limit $S_{\mathrm{lim}}$ are 
considered: 
\begin{equation}
\delta F_{\mathrm{lim}} (\lambda_i, S_{lim}) = \left(\frac{1}{\Omega_c}\right)
\sum\limits_{S_j<{S_{\mathrm{lim}}}} S_j^2(\lambda_i)
\label{eq:conf_sum}
\end{equation} 

\noindent Note that \dflim\, is instrument-dependent only through the actual 
sensitivity limit, and other characteristics of the instrument,
like spatial resolution, are not taken into account in the calculation 
of \dflim. These properties are considered in the calculation of the
confusion noise. \dfnull\, is a fully instrument-independent
quantity, and both \dfnull\, and \dflim\, depend strongly on the
actual asteroid model.  

\smallskip
\noindent{\bf Confusion noise}:
Throughout this paper we assume that the \emph{local} spatial 
distribution of the asteroids is Poissonian, i.e.~the same fluctuation
power can be used to calculate the confusion noise at any spatial frequency,
independent of the instrument.  
Thus confusion noise can be calculated from the fluctuation powers as:
\begin{equation}
\sigma_0(\lambda_i, \Omega_p) = \Big( \Omega_p \cdot \delta F_0(\lambda_i) 
    \Big) ^{1\over{2}} 
\label{eq:sigma0}
\end{equation}   
\vspace{-0.6cm}
\begin{equation}
\sigma_{\mathrm{lim}}(\lambda_i,S_{\mathrm{lim}}, \Omega_p) = 
    \Big(  \Omega_p \cdot 
      \delta F_{\mathrm{lim}}(\lambda_i, S_{\mathrm{lim}}) \Big) ^{1\over{2}}
\label{eq:sigmalim}
\end{equation}    
for the 'full' and sensitivity limit dependent confusion noise values,
respectively. $\Omega_{\rm p}$ is the effective solid angle of 
the detector, which is \emph{not} necessarily the physical size of the
actual pixel/aperture,  and the confusion
noise applicable for detection limits of point sources depends
on the point source flux extraction/reconstruction method as well, 
\citep[see e.g.][]{Kiss2005}.
"Best estimates" of effective solid angles of various instruments
and filters can be found at our 
webpage\footnote{http://kisag.konkoly.hu/solarsystem/irsam.html}.

Due to their relatively low number density, asteroids in our model
limit the detectability of point sources through the 
\emph{photometric}, rather than the number density limit
\citep[see][for a detailed introduction]{Lagache2003}.
\cnfull~and \cnlim~are \emph{lower limits}, since 
there is an unknown contribution of small (fainter) asteroids, which is
not considered here (see Sect.~4 for a discussion).

In the following, unless otherwise quoted, we refer to all these
quantities as specific for a given wavelength or photometric band and 
instrument, and therefore the wavelength, sensitivity limit and spatial 
resolution (effective detector solid angle) dependences are not marked. 
 

\section{Results}
\label{sect:results}

\subsection{General results}
\label{sect:generalres}

The positions of the asteroids in the original SAM model were
integrated in the time span January~1, 2000 to December~31, 2012. 
At each date a spectral energy distribution was assigned to each asteroid using the
STM, as described in Sect.~\ref{sect:tbr}. 
Fluctuation powers and number counts have been derived from this 
database, which is publicly available at the 
URL:"{\sl http://kisag.konkoly.hu/solarsystem/irsam.html}". 
The web-interface is described in more detail in Appendix~A.

For some specific instruments and time spans we constructed 
\dfnull, \dflim, \ntot, and \nlim\, maps to characterize 
the impact of main belt asteroids for the selected instruments. 
An example is shown in Fig.~\ref{fig:bigmap}.

The fluctuation power and number count maps at different
wavelengths (and the same date) show a very similar morphology.
Considering the celestial structure, the main characteristics 
are the following (presented in the \emph{ecliptic} coordinate system):
\begin{itemize}

\item The \dfnull~ and \ntot~ distributions (and so \dflim~ and \nlim)
are symmetric in ecliptic latitude ($\beta$), and show a maximum
at the ecliptic plane at a specific ecliptic longitude ($\lambda$);
\dfnull, \dflim, \ntot~and \nlim~show a similar morphology in
celestial distribution.

\item  The maximum extension
in $\beta$ is at the anti-solar point, where fluctuation power 
isocontours form a "bulge", while the minimum is at
the celestial position of the Sun. When comparing different dates, 
the bulge around the anti-solar point moves along the ecliptic
as Earth revolves around the Sun, and so does the minimum. 

\item If maps presented in the ecliptic coordinate systems 
($\lambda$--$\beta$ maps) are transformed to helioecliptic
coordinate system ([$\lambda - \lambda_{\odot}$]--$\beta$, 
where $\lambda_{\odot}$ is the ecliptic longitude of the Sun),
then the maps of different dates are rather similar. The differences
are about one order lower than the median values of the different maps. 
Therefore it is possible to create time-independent average maps 
for a specific instrument setup, which can give 'average' 
fluctuation power and number count estimates, without the need
to specify the exact day of the observation.

\end{itemize}



The main component of the asteroid fluctuation power (or confusion noise)
as well as the number counts can be well represented by a map,
which is constant in the helioecliptic 
coordinate system. This kind of map can serve as a good guideline
to characterize the expected impact of main belt asteroids 
for the infrared and submillimetre measurements of a specific
instrument, if the exact date of the observation is not known. 
We present these maps for a handful of instruments in 
Appendix~B, and they are also available in FITS format at our 
webpage$^4$.

For medium ecliptic latitudes the asteroidal sky changes 
in a timescale of a few weeks, with an actual value 
depending on the instrument (wavelength and sensitivity limit)
and sky position. Close to the ecliptic plane the amplitude of 
these temporal changes are less pronounced, and there are practically
no changes at high ecliptic latitudes since asteroids are 
present at these places only sporadically.


\subsection{The importance of main belt asteroids for specific instruments}
\label{sect:res-cnoise}

Asteroids may
affect IR and submillimetre observations in two ways: they can increase
the fluctuation power (and hence the confusion noise) level, 
and contribute to the number count of point sources.
In our investigated wavelength regime the main sources of 
confusion noise are the extragalactic background and 
the Galactic cirrus emission. The strength of the cirrus
emission and confusion noise changes rapidly from place to place
in the sky and is below the extragalatic confusion noise level
in the best 'cosmological' windows for most of the space 
IR instruments \citep[see e.g.][]{Kiss2005}. 

Being constant and present in any direction, extragalactic background
fluctuations represent a {\it minimum} value for the confusion noise. 
Therefore we used the respective extragalactic
fluctuation powers to judge, whether asteroid confusion noise
has to be considered for a specific instrument. In 
the calculations presented in Table~\ref{table:cncomp},
we considered an instrument as "affected", if the asteroid confusion 
noise level was
at least half of the extragalactic background component
\citep[as calculated by][]{Lagache2003,Lagache2004}.  

In the case of infrared space instruments there is always a 
solar elongation constraint for the actually observable part 
of the sky, e.g. \lele{60\degr}{$\lambda-\lambda_{\odot}$}{120\degr}
for the Herschel Space Telescope, 
\lele{85\degr}{$\lambda-\lambda_{\odot}$}{120\degr}
for the Spitzer Space Telescope and 
\lele{89\degr}{$\lambda-\lambda_{\odot}$}{91\degr} for Akari.
Instruments of these spacecrafts can never look at or close to the 
anti-solar point, where the highest asteroid fluctuation power is 
expected. 

As a general result of the comparison of different wavelengths, 
instruments working in the far-infrared regime
are not affected, while instrument working in the 
$5~\mu\mathrm{m}\le\lambda_i\le30~\mu\mathrm{m}$ domain can be severely
affected by asteroid confusion noise, at least along or close to 
the ecliptic. Although far-infrared instruments are not affected by the
asteroid confusion noise, asteroids above the 
detection limit may have a considerable impact on the source counts.
This contribution can be best estimated for a specific measurement
(instrument, date, sky position) by the web-interface of our 
infrared asteroid model.

\section{Discussion}


The SAM is limited to asteroids with a lower limit in size of 
$1~\mathrm{km}$ in diameter. There is, certainly, a population of asteroids
with sizes below this limit. Since the confusion noise is calculated
using Eq.~\ref{eq:conf_sum}, the impact of very small boides (a few
hundred meters in diameter and below) is minor to the confusion noise.

To test the effect of small ($D<1~\mathrm{km}$) asteroids on the fluctuation 
power and confusion noise we first created a size-frequency distribution;
for asteroids of $D>1~\mathrm{km}$ we used the number count values
estimated by eq.~3 in \citet{Tedesco2005}, while for $D\le1~\mathrm{km}$ 
the interpolated values of \citet{Belton1992} were used. This 
SFD is a combination of the 'SAM' and 'Galileo team' data points in
fig.~4 in \citet{Tedesco2005}. A simple relationship of 
$S(\lambda_i)\propto D^2$ was assumed between the flux density 
of an asteroid's thermal emission at a specific wavelength 
$S(\lambda_i)$ and its diameter $D$. Then the specific fluctuation 
power $\delta F(D)$ was calculated for a $D\pm\Delta D$ interval,
using Eq.~\ref{eq:dF0}.  
\begin{figure}[!ht]
\includegraphics[width=7.8cm]{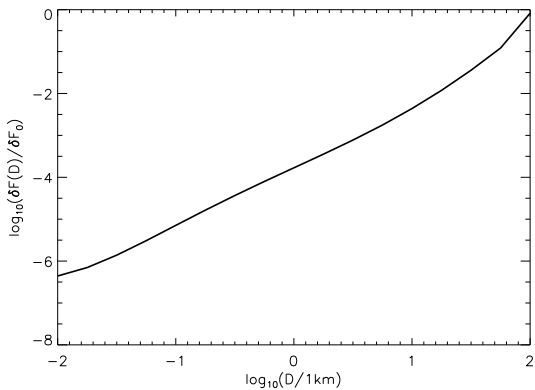}
\caption[]{Relative contribution of asteroids with a specific size
$D$ to the asteroid fluctuation power, using the size distribution and
the simple flux model as described in Sect.~4.}
\label{fig:below1km}
\end{figure}
It is clear from Fig.~\ref{fig:below1km} that the fluctuation power
is dominated by the bright and large asteroids, and that small 
($D\le1~\mathrm{km}$) minor planets have a negligible contribution to the 
fluctuation power. 
{ Presently orbits for almost 400\,000 asteroids are known,
i.e., more than 20\% of the 1.9$\cdot$10$^6$ SAM asteroid sample 
(see Sect.~1), but only about 0.1\% ($\sim$2000 asteroids) 
have known sizes and albedos which are the crucial parameters
for the confusion noise estimates. The calculated power
fluctuations are therefore dominated by asteroids which have
been extrapolated in the SAM via size-frequency distributions.
However, knowing only 0.1\% is
far from being statistically relevant and in some parts 
of the sky bright (i.e.~large) asteroids are completely missing. 
This implies that confusion noise estimates cannot rely on the known 
asteroid sample alone and the application of the SAM model for
confusion noise calculations is necessary. 

On the other hand, fluctuation power and confusion noise are dominated 
by the largest/brightest asteroids in the actual field, even if the
known ones are missing. As suggested by Fig.~\ref{fig:below1km}
the contribution of $D\le1~\mathrm{km}$ asteroids remains 
negligible, and a further extension of the SAM to smaller diameters would 
not improve the accuracy of the  
confusion noise calculations significantly.  
}

Asteroids further out in the Solar System 
(e.g. trans-Neptunian objects) will not contribute considerably to 
the confusion noise or number counts, 
as deduced from their currently known size distribution
in Appendix~C.  

As discussed in detail in Appendix~D, 
SAM asteroids contribute to the absolute brightness of the 
zodiacal emission in a negligible level, however, a set of these
asteroids may be the dominant sources of zodiacal emission 
fluctuation power at some specific wavelengths and sky regions.

\begin{acknowledgements}
\sloppy\sloppy
This research was supported by the European Space Agency (ESA) and 
by the Hungarian Space Office via
the PECS programme (contract no. 98011). Cs.\,K. and P.\,\'A. acknowledge the 
support of the Hungarian Reserch Fund (OTKA\,K62304).
\end{acknowledgements}

{}

\clearpage
\newpage

\begin{figure*}[!ht]
\begin{center}
\includegraphics[width=16cm]{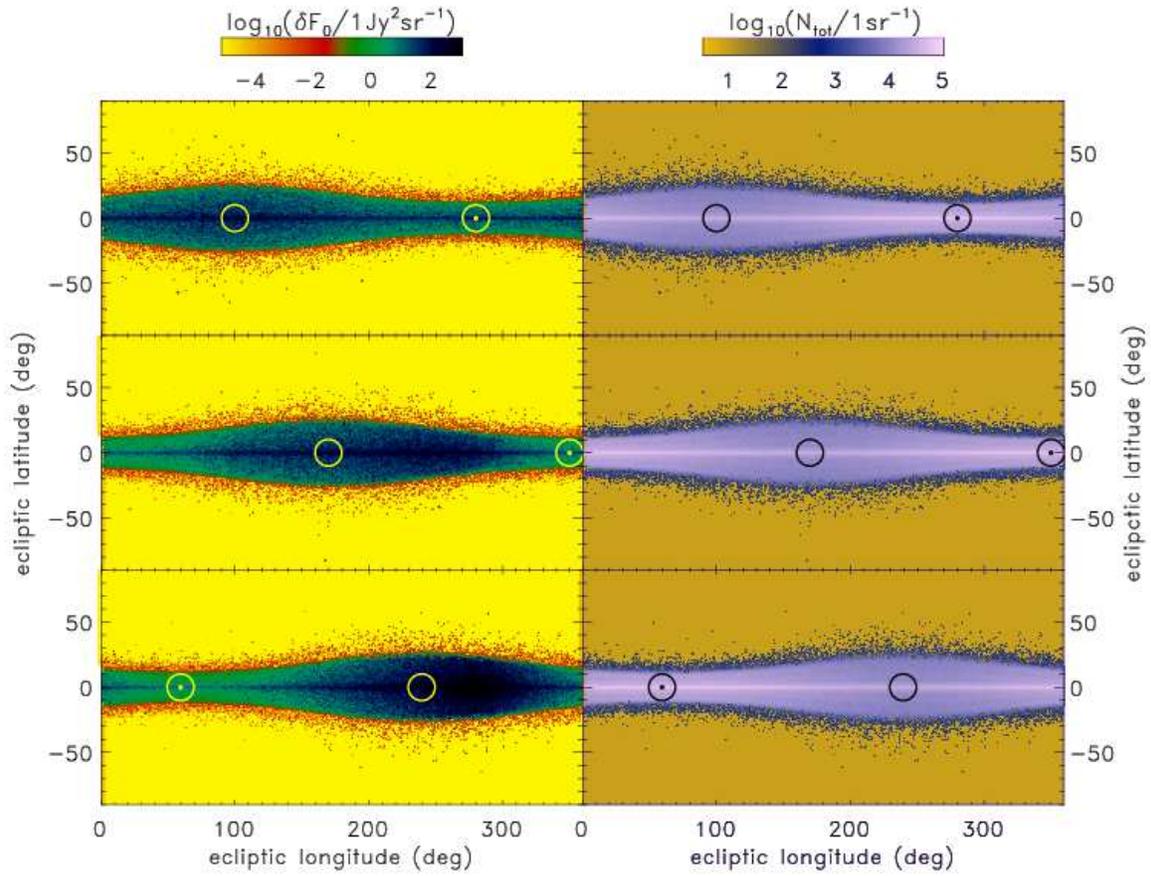}
\end{center}
\caption[]{Expected distribution of fluctuation power 
(\dfnull, left column) and the total number of asteroids
(\ntot, right column)
over the sky of SAM asteroids at $20~\mu\mathrm{m}$ on -- from top to
bottom -- January~1, March~10 and May~20, 2008 (ecliptic coordinate
system). The position of the solar and anti-solar points are marked with
"$\bigodot$" and "$\bf\bigcirc$" symbols, respectively.}
\label{fig:bigmap}
\end{figure*}
\clearpage

\begin{table*}
\begin{tabular}[]{lrrrrc}
\hline
Instrument & $\lambda_i$ & 
  \multicolumn{2}{c}{\ensuremath{\log_{10}(\delta F_0/\mathrm{Jy}^2~\mathrm{sr}^{-1})}} 
    & $\lambda-\lambda_{\odot}$ & $|{\Delta\beta}|$ \\
  & ($\mu\mathrm{m}$) & (EGB) &  (MBA) & (deg) & (deg) \\   
\hline
Spitzer/IRAC	              &   8 & 1.49 &  2.11 & 120 &  20 \\
Akari/IRC                     &   9 & 1.47 &  1.21 &  90 &  13 \\
Akari/IRC  	              &  18 & 1.39 &  2.23 &  90 &  16 \\ 
Spitzer/MIPS	              &  24 & 2.04 &  2.70 & 120 &  21 \\
\hline
Akari/FIS                     &  65 & 3.47 &  1.34 &   90 &  -- \\
Spitzer/MIPS \& Herschel/PACS &  70 & 3.69 &  1.62 &  120 &  -- \\
Akair/FIS                     &  90 & 4.06 &  0.98 &   90 &  -- \\
Herschel/PACS	              & 100 & 4.42 &  0.95 &  120 &  -- \\
Akari/FIS                     & 140 & 4.42 &  0.41 &   90 &  -- \\
Akari/FIS                     & 160 & 4.43 &  0.23 &   90 &  -- \\ 
Spitzer/MIPS \& Herschel/PACS & 160 & 4.43 &  0.36 &  120 &  -- \\
Herschel/SPIRE 	              & 250 & 4.29 & -0.21 &  120 &  -- \\
Herschel/SPIRE 	              & 350 & 3.88 & -0.59 &  120 &  -- \\
Herschel/SPIRE 	              & 500 & 3.27 & -1.64 &  120 &  -- \\
\hline 
\end{tabular}
\caption[]{
Comparison of the extragalactic and asteroid full
fluctuation power levels at some representative wavelengths of 
infrared and submillimetre space instruments. Asteroid fluctuation power was derived
at the specified $\lambda$--$\lambda_{\odot}$ helioecliptic longitude, and
at the ecliptic plane. The columns of the table are: 
1) name of the instrument;
2) wavelength;
3) extragalactic background fluctuation power (EGB);
4) asteroid fluctuation power (MBA);
5) solar elongation used in the MBA fluctuation power calibration, 
in accordance with the visibility constraints of the instrument;
6) Ecliptic latitude range above/below the ecliptic plane, where a 
specific instrument is considered as 'affected' by asteroid confusion
noise. Missing values indicate that the instrument is
\emph{not} affected. }
\label{table:cncomp}
\end{table*}

\end{document}